\documentclass[twocolumn,superscriptaddress,prl]{revtex4-1}% Physical Review B

\usepackage{amsmath,amssymb,graphicx}
\usepackage{float}
\usepackage{latexsym} 
\usepackage{amsmath,amsthm}
\usepackage{ifpdf}
\usepackage{epstopdf}
\usepackage{dcolumn}% Align table columns on decimal point
\usepackage{bm}% bold math
\usepackage{braket}
\usepackage{color}
\usepackage{graphicx}
\DeclareMathOperator{\sech}{sech}

\newcommand{\hatF}{\hat{\phi}}%
\newcommand{\hatFd}{\hat{\phi}^\dag}%
\begin{document}

\def \beq{\begin{equation}}
\def \eeq{\end{equation}}
\def \bea{\begin{eqnarray}}
\def \eea{\end{eqnarray}}
\def \bes{\begin{split}}
\def \ees{\end{split}}
\def \besu{\begin{subequations}}
\def \esu{\end{subequations}}
\def \bea{\begin{align}}
\def \eal{\end{align}}
\def \bem{\begin{displaymath}}
\def \eem{\end{displaymath}}
\def \P{\Psi}
\def \Pd{|\Psi(\boldsymbol{r})|}
\def \Pds{|\Psi^{\ast}(\boldsymbol{r})|}
\def \Po{\overline{\Psi}}
\def \bs{\boldsymbol}
\def \dert{\frac{d}{dt}}
\def \k{\ket}
\def \b{\bra}
\def \bl{\bar{\boldsymbol{l}}}
\def \s{\sum_{n=1}^\infty}
\def \so{\sum_{n=0}^\infty}
\def \u{\underline}
\def \derfua{\frac{\delta}{\delta \u \alpha}}
\def \derfub{\frac{\delta}{\delta \u \beta}}
\def \hatL{\hat \lambda(\u \alpha,\u \beta)}
\title{Quantum Soliton Evaporation}
\author{Leone Di Mauro Villari}
\affiliation{University Sapienza, Department of Physics, Piazzale Aldo Moro 5 00185, Rome, Italy}
\affiliation{Institute for Complex Systems, National Research Council, (ISC-CNR), Via dei Taurini 19, 00185, Rome (IT) }
\author{Ewan M. Wright}
\affiliation{College of Optical Sciences, University of Arizona, Tucson, AZ 85721,USA}
\affiliation{Institute of Photonics and Quantum Sciences, Heriot-Watt University, Edinburgh EH14 4AS, UK}
\author{Fabio Biancalana}
\affiliation{School of Engineering and Physical Sciences, Heriot-Watt University, EH14 4AS Edinburgh, United Kingdom}
\author{Claudio Conti}
\affiliation{University Sapienza, Department of Physics, Piazzale Aldo Moro 5 00185, Rome, Italy}
\affiliation{Institute for Complex Systems, National Research Council, (ISC-CNR), Via dei Taurini 19, 00185, Rome (IT) }

\begin{abstract}
We have very little experience of the quantum dynamics of the ubiquitous nonlinear waves. Observed phenomena in high energy physics are perturbations to linear waves, and classical nonlinear waves, like solitons, are barely affected by quantum effects.  We know that solitons, immutable in classical physics, exhibit collapse and revivals according to quantum mechanics. However this effect is very weak and has never been observed experimentally.
By predicting black hole evaporation Hawking first introduced a distinctly quantum effect in nonlinear gravitational physics.
Here we show the existence of a general and universal quantum process whereby
a soliton emits quantum radiation with a specific frequency content, and a temperature given by the number of quanta, the soliton Schwarzschild radius, and the amount of nonlinearity, in a precise and surprisingly simple way.
This result may ultimately lead to the first experimental evidence of genuine quantum black hole evaporation. 
In addition, our results show that black hole radiation occurs in a fully quantised theory, at variance with the common approach based on quantum field theory in a curved background; this may provide insights into quantum gravity theories. Our findings also have relevance to the entire field of nonlinear waves, including cold atomic gases and extreme phenomena such as shocks and rogue-waves. Finally, the predicted effect may potentially be exploited for novel tunable quantum light sources.
\end{abstract}

\maketitle
\begin{figure}
\includegraphics[width=1\columnwidth]{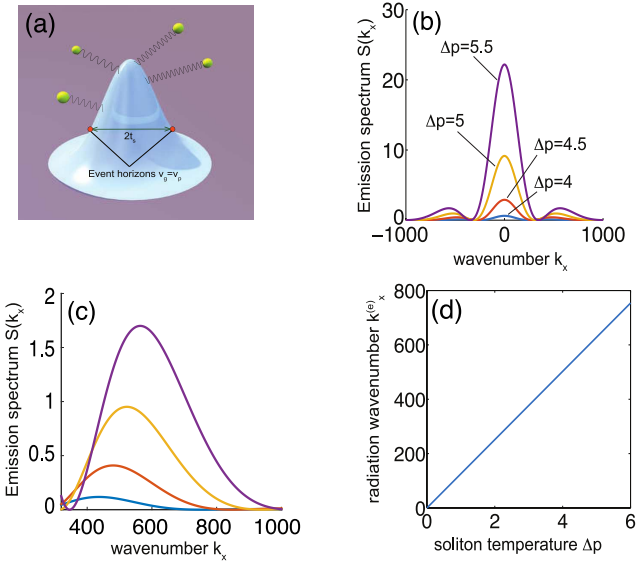}
\caption{(a) Sketch of the soliton event horizon. Here we consider the case of temporal solitons at a fixed instant with intensity profile $I(t)=I_0$sech$^2(t/t_N)$.
(b) Emission spectrum according to Eq. (\ref{spec}) at a fixed normalised time $\tau=1$ and versus $\Delta p$ in units of $\kappa n_0$. (c) The quantum contribution to emission spectrum obtained as the difference between the full spectrum and the classical one. (d) Position of the maximum of the lateral peak $k_x^{(e)}$ versus the soliton temperature $\Delta p$. \label{SEHorizon}}
%\end{center}
\end{figure}

Solitons are localized analytical solutions of integrable nonlinear partial differential equation, such as the Korteweg-De Vries equation or the nonlinear  Schr\"{o}dinger equation (NLSE) \cite{DrazinBook,zach:72}. Solitons are ubiquitous, and occur in many different contexts, including light \cite{agr, Kiv}, ultracold gases \cite{bs}, sound and water waves \cite{k-m}, and even in the brain synapses or in DNA dynamics \cite{th, Bertrand, zt}.
As noted a long time ago by Abdus Salam in a largely overlooked paper \cite{salam}, black holes (BH) are also a class of solitons, because they are analytical solutions of the highly nonlinear Einstein-Hilbert equations, which are solvable via the inverse scattering transform technique \cite{belinski,b-z}.
The event horizon (EH) in a BH is the classical ``point of no return'' for photons and other particles. However, according to quantum mechanics, photons escape the BH \cite{H,P}. As a photon escapes, an amount of energy $\Delta E$ is lost, and the BH mass/energy reduces generating the evaporation.
The uncertainty principle (UP) determines $\Delta E$
in terms of the Schwarzschild radius $r_s$,
\begin{equation}
\frac{r_s}{c} \Delta E \simeq \frac{\hbar}{2}\text{,}
\label{UP}
\end{equation}
with $\hbar$ the reduced Planck constant and $c$ the vacuum light velocity.
The photon emission exhibits a black body spectrum at the Hawking temperature 
\begin{equation}
T = \frac{\hbar c^3}{8\pi k_B GM}=\frac{\hbar c}{4 \pi k_B r_s}\text{,}
\label{HawkT}
\end{equation}
being $M$ the BH mass, $G$ the gravitational constant,  and $k_B$ the Boltzmann constant \cite{cg,s,tc,Steinhauer2016}.
We show here that this emission process also occurs for \emph{integrable} NLSE quantum solitons. This is surprising since integrable solitons are classically the most robust objects in nonlinear physics, their dynamics being protected by an infinite number of conservation laws \cite{DrazinBook, zach:72}. 
With specific reference to temporal solitons, for example those observed in optical fibres \cite{Kiv,agr,ls,sf,yl,lr}, the EH is determined by that instant $t_s$ within the soliton profile at which phase velocity $v_p$ is equal to group velocity $v_g$ (Figure \ref{SEHorizon}) \cite{lp}.  In a nonlinear optical Kerr medium, the refractive index $n_r=n_{r0}+n_2 I$ is a function of the intensity $I$, with $n_{r0}$ the linear bulk index and $n_2$ the Kerr coefficient. The phase velocity is $v_p=c/n_r(I)$ and the {\it soliton event horizon} (SEH) is given by
\begin{equation}
n_{r0}+n_2 I(t_s)=c/v_g
\label{horizondefinition}
\end{equation}
As illustrated in Fig.~\ref{SEHorizon}(a), Eq. (\ref{horizondefinition}) defines $t_s$.
\noindent From $t_s$ we can find the soliton {\it black body temperature}, in full analogy with Eq. (\ref{HawkT})
\begin{equation}
T = \frac{\hbar }{4\pi k_B t_s}\text{,}
\label{HT}
\end{equation}
which provides the spectral position of the peak quantum emission process for the soliton. Similar arguments also hold true for other kinds of solitons as, for example, those reported in Bose-condensed gases \cite{bs}.

In this paper we validate these heuristic arguments and elucidate the underlying physics using the exact quantum soliton theory.  In particular, we show that solitons emit radiation at a characteristic wavevector and frequency, with resulting energy loss. This provides proof of the quantum soliton evaporation from first principles.

We use the exact second quantisation form of the NLSE \cite{Lai:89, Lai2:89, yao:95}.  By adopting the units in \cite{Lai:89, Lai2:89, yao:95}, denoted hereafter as the {\it quantum soliton units}, the quantum NLSE (QNLSE) reads
\begin{equation}
i\hatF_\tau(\xi,\tau)+\hatF_{\xi\xi}(\xi,\tau) + 2\kappa \hatFd(\xi,\tau)\hatF(\xi,\tau)\hatF(\xi,\tau)=0.
\label{QNLSE}
\end{equation}
In Eq. (\ref{QNLSE}) the coefficient $\kappa >0$ measures the strength of the self-focusing nonlinearity. The Bethe ansatz and quantum inverse scattering method \cite{creamer:79,Lai2:89} furnish the fundamental soliton as a superposition of the states $\ket{n,p}$, with weights $g_n(p)$
\beq
\k {\psi_s,\tau} = a_0\k 0 + \s a_n \int dp\,g_n(p) e^{-iE_{n,p} \tau} \k{n,p}  .
\eeq
Here $\Delta p$ is the variance of the distribution $g_n(p)$, and the classical soliton corresponds to $\Delta p=0$. As shown below, $\Delta p$ plays the role of the {\it soliton temperature}, and the spectral emission power is proportional to $\Delta p$. 
In the most general case, the quantum expectation value of the field operator $\hatF(\xi)$ with respect to the state $\k \psi$ is given by the average of a set of classical soliton solutions with different group and phase velocities. The quantum soliton therefore displays phase diffusion and wave-packet spreading \cite{Lai2:89}. \\

In dimensional units, and with reference to an optical fiber \cite{agr}, we have the following expression for the soliton width (calculated as the standard deviation) during propagation along the $z$-axis:
\begin{equation}
\langle\Delta t^2\rangle\approx \Delta t_0^2 + \frac{\hbar^2 \omega_0^2}{\Delta t_0^2 \mathcal E^2}\beta_2^2 z^2  ,
\label{qfp2}
\end{equation}
where $\beta_2<0$ and $\beta_1$ are the dispersion coefficients \cite{agr}, $\omega_0$ is the optical angular frequency, $\mathcal{E}$ is the soliton energy, and $\Delta t_0$ is the initial soliton width. 

Equation (\ref{qfp2}) gives the spreading of the soliton width during propagation. For $\hbar \to 0$ the $z-$dependent part of (\ref{qfp2}) goes to zero,
and this corresponds to the fact that the classical soliton does not disperse in absence of quantum effects. 
The same occurs when the number of quanta $n_0=\mathcal{E}/\hbar \omega_0$ in the soliton goes to infinity.  In contrast, in the quantum regime the soliton exhibits collapse and revivals \cite{w,g-d} while emitting its energy content.

We next show that the quantum collapse is accompanied by emission of radiation in the form of spectral sidebands. 
Using the exact quantum theory (details in the Supplemental Material) we obtained the spectrum $S(k_\xi,\tau)$ of the field operator $\hatF(\xi)$ as 
\begin{widetext}
\begin{equation}
S(k_\xi,\tau)=|\braket{\psi_s|\hat a|\psi_s}|^2 \approx  \frac{1}{2 \bar p^2 n_0^2\tau^2} \sech^2\left(\frac{\pi k_\xi}{\kappa n_0}\right)e^{-\frac{2\bar p^2}{\Delta p^2}-\frac{2k^2_\xi}{n_0^2 \Delta p^2}}
\left[\cosh\left(\frac{2}{\Delta p^2}\frac{k_\xi}{n_0} \bar p\right) + \cos\left(\bar p k_\xi \tau\right)\right]\text{.}
\label{spec}
\end{equation}
\end{widetext}
Equation (\ref{spec}) is the quantum counterpart of the modulus squared of the Fourier transform of the classical field, and is the expectation of the annihilation operator $\hat a$, which is in turn the Fourier transform of the field operator $\hat \phi$, and gives the spectrum in terms of the momentum cut-off $\bar p=\omega_0/c$ \cite{DH}.

Figures \ref{SEHorizon}(b,c) show the shape of the spectrum according to Eq.~(\ref{spec}) for various powers, and reveal that, in addition to the dominant low frequency classical spectrum represented by the central peak in Fig. \ref{SEHorizon}(b), one finds a quantum contribution peaked at a characteristic frequency $k_x^{(e)}$ shown in Fig. \ref{SEHorizon}(d). The lateral spectral tails in Fig. \ref{SEHorizon}(c) correspond to the spontaneous emission from the soliton and the resulting evaporation. 

From Eq. (\ref{spec}) we may compute the relation between the emission wavenumber $k_\xi^{(e)}$ and $\Delta p$. 
For quantum solitons with $n_0>> n_0 \kappa>>1$, and considering the emission on the timescale of the soliton collapse, i.e. $\tau_1=1/(n_0\kappa \Delta p)$ \cite{g-d}, we have from Eq. (\ref{spec})
\begin{equation}\label{S9}
S(k_\xi ,\tau_1)\approx \frac{2}{n_0^2\tau^2} \sech^2\Bigl(\frac{\pi k_\xi}{n_0\kappa}\Bigl)
\Bigl[1+\cos\Bigl(\frac{\bar p k_\xi}{n_0 \kappa\Delta p}\Bigl)\Bigl]\text{.}
\end{equation}
The corresponding peak emission then ocurs at
\begin{equation}
k_\xi^{(e)}= \pi n_0 \kappa \frac{\Delta{p}}{\bar p}  .  % \,\,\,\ n \in \mathbf{Z}\text{.}
\label{exactHT}
\end{equation}
Equation~\eqref{exactHT} shows the linear dependence of the emission peak with $\Delta p$ as seen in Fig. \ref{SEHorizon}(d), and anticipated in Eq. (\ref{HT}).  Thus $\Delta p$ plays the role of temperature in quantum soliton units.

The relation alluded to above between the quantum soliton collapse and the emission peaks calls for elaboration.  One of the present authors (EMW) first calculated the collapse and revival of a quantum soliton under the approximation that the center-of-mass (COM) of the underlying $N=1$ soliton was localized at a fixed but arbitrary position, this approach showing no emission peaks \cite{w}.  This calculation was extended in Ref. \cite{g-d} to allow for quantum diffusion of the soliton COM, and that approach was utilized here in Eq. (\ref{S9}) to reveal the emission peaks.  This establishes the connection between the quantum diffusion and the emission.   Intuitively one may then view the quantum diffusion as due to a series of emission events from the underlying $N=1$ soliton, the COM recoiling after each such event.  This notion of emission from the underlying soliton is contrary to classical theory in which an energy gap ensures stability.  However, exact diagonalization of the QNLSE for large particle numbers $N$ shows that the energy gap between the ground and first excited states scales as $1/N$, and the ground state becomes vulnerable to coupling to a set of quasi-degenerate translational states representing the soliton COM motion \cite{KanSaiUed06}.  This suggests an open systems viewpoint in which the evolution of the quantum soliton is envisaged as the interaction of the underlying $N=1$ soliton with a reservoir comprising the COM motional states responsible for the quantum diffusion described in Eq. (\ref{qfp2}).  From this open system viewpoint the quantum collapse and concomitant emission and evaporation result from the decoherence and loss arising from the coupling between the underlying soliton and its COM motion \cite{MB}.

A further remark is in order: We calculated the spectrum in Eq. (\ref{spec}) as the square modulus of the quantum expectation value of the annihilation operator $|\langle a \rangle|^2$. This is a good approximation to the spectrum for quantum states close to a coherent state. Alternatively one could calculate the spectrum as $\langle a^\dagger a\rangle$.  We verified numerically that these two choices of spectrum do not differ markedly in our simulations, and that the spectral peak in Eq.(\ref{spec}) is always present. The relative insensitivity of the spectrum to the particular choice may be ascribed to the specific features of the quantum solitons, which, even for a finite number of photons, seem to preserve a nearly coherent state description: This is consistent with the open system viewpoint above in that it is well known that coherent states are maximally stable with respect to interaction with a reservoir \cite{ZurHabPaz93}.  Interesting though these issues are, we relegate a detailed description of them to a future publication and here concentrate on the emission.

To further support our analysis, we resort to the numerical solution of the full quantum model. The QNLSE is an operator equation and cannot be directly solved numerically. Phase space methods map the QNLSE to equivalent stochastic differential equations (SDEs) \cite{G-Z, Gardiner}. We use the positive P-representation that transforms the Heisenberg equations of motion into a Fokker-Planck equation \cite{Glauber, G-Z, DH} which corresponds to a It\^o set of differential equation \cite{Gardiner}. The quantum field is then represented by a stochastic field, and ensemble averages built from realizations of the stochastic classical field provide the observable quantum averages. 
The It\^o set of differential equation is solved numerically by a second-order pseudo-spectral stochastic Runge-Kutta algorithm \cite{K2}. 

\begin{figure}
%\begin{figure}[H]
%\begin{center}
\includegraphics[width=\columnwidth]{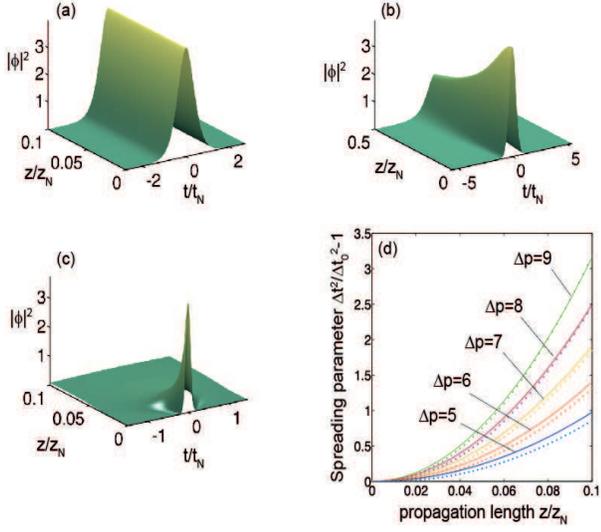}
\caption{Square modulus of the quantum-average of soliton field  s functions of both propagation distance, in units of the dispersion length $z_N=t_N^2/\beta_2$, and time in (a) the classical, and quantum regimes (b) $\Delta p=2\kappa n_0$, and (c) $\Delta p=6\kappa n_0$. Plot (d) shows the variance of the quantum soliton with respect to the classical soliton at various $\Delta p$ in units of $\kappa n_0$.
}
\label{evaporation}
%\end{center}
\end{figure}

We first numerically validate the quantum spreading in Eq. (\ref{qfp2}) by studying the transition from the evolution of the invariant classical soliton to the evaporating quantum soliton.  Figure \ref{evaporation} shows the dynamics of the quantum soliton with increasing $\Delta p$. As in the BH case, zero temperature $\Delta p=0$ corresponds to a classical theory and absence of spreading as shown in Fig. \ref{evaporation}(a). For $\Delta p>0$ Figs. \ref{evaporation}(b,c) very clearly show the quantum spreading during the nonlinear propagation, the amount of quantum spreading increasing with $\Delta p$. 

When $\Delta p<<\kappa n_0$ and $n_0>>1$ the center-of-mass of the soliton is localised and the quantum fluctuations are irrelevant \cite{Lai2:89}.  In previous experiments this condition is always satisfied, and the quantum spreading is negligible and has never been observed \cite{D-S}. Moreover, it is possible to show \cite{Lai:89,Lai2:89} that the wave-packet spreading is smaller compared to the phase diffusion effect and, as a consequence, it is very difficult to observe signatures of the quantum spreading.  Experiments have only reported evidence of the phase diffusion \cite{D-S}. In Fig. \ref{evaporation}(d) we show the calculated variance of the stochastic field and find very good agreement between theory and simulations. As the quantum fluctuations are irrelevant in the limit $\Delta p<<\kappa n_0$ the quantitative comparison with the theory improves when $\Delta p$ increases. From the numerical simulations, it is possible to estimate the normalised collapse  distance $z_1/z_N$ defined by the distance over which the width of the soliton is doubled. We have  $z_1/z_N  \approx \frac{\kappa n_0}{4\Delta p  }$ which corresponds to the collapse  time $\tau_1=\frac{1}{n_0\kappa}$  in quantum soliton units.

\begin{figure}
%\begin{figure}[H]
%\begin{center}
\includegraphics[width=1 \columnwidth] {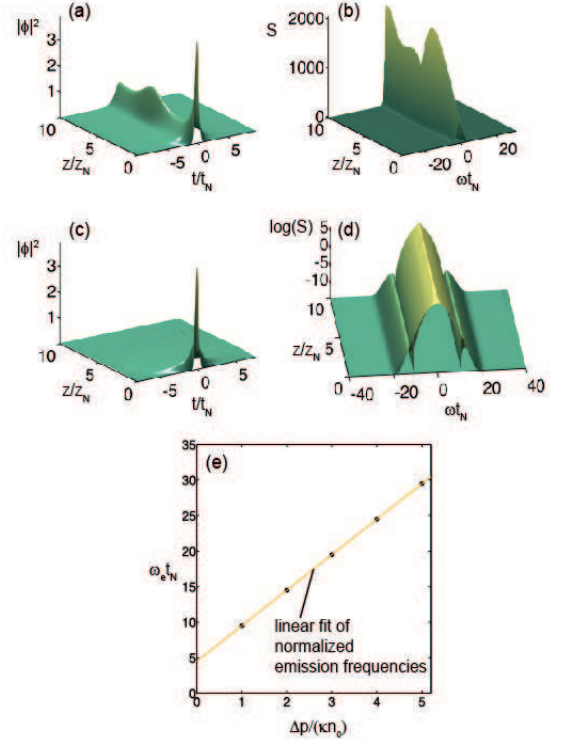}
\caption{(a) Propagation dynamics in the regime of collapse and revival dynamics, and (b) corresponding spectrum.  (c) Propagation dynamics in the evaporation regime, and (d) corresponding spectrum. (e) Peak spectral position versus soliton temperature.
\label{spectrum}}
%\end{center}
\end{figure}

In contrast to the quantum soliton spreading, the emission spectrum of quantum origin can be readily visible in the experiments by direct spectral measurements.
In our SDE simulations we have clear evidence of the spectral emission accompanying quantum spreading:
When increasing $\Delta p$ we observe a weak emission process together with the collapse and revivals (Fig. \ref{spectrum}(a)), as originally studied in Refs. \cite{w,g-d}. Here we show that these processes also include the periodic formation of emission sidebands. For the specific case of collapse and revival (Fig \ref{spectrum}(b)), very long propagation distances are needed in the nearly classical regime. This breathing regime is found for $1 \leqslant \Delta p \kappa / n_0 < 1.5$.

Our key result is obtained in the strongly quantum regime, when the breathing dynamics have abated and the soliton evaporation spectrum remains constant after the collapse (Fig. \ref{spectrum}(c,d)). Indeed after the collapse all the initial energy content is emitted in the black-body spectrum that remains unaltered upon further propagation. 
The SDE confirms the analytical result and the peak-emission wave-number is directly proportional to the value of the momentum fluctuation $\Delta p \kappa  / n_0$ as shown in Fig.~\ref{spectrum}(e).
Writing Eq. (\ref{exactHT}) dimensional units, and for the case of solitons in an optical fiber,  we obtain for the peak frequency $\omega_e$ of the emitted spectrum (corresponding to $k_\xi^{(e)}$ in quantum soliton units)
\begin{equation}
\omega_e=\frac{\pi \hbar\omega_0   \beta_1 n_2 }{A_{eff}|\beta_2| n_0 \Delta t_0} ,
\label{omegae}
\end{equation}
where $A_{eff}$ is the effective area \cite{agr}.
\begin{table}
%\begin{table}[H]
%\begin{center}
\begin{tabular}{|c|c|}
\hline $\Delta t_0$ $(fs)$ & $\omega_e$ $(Hz)$\\
\hline $10^3$ & 1$0^8$\\
\hline $10^2$ & $10^9$\\
\hline $10$ & $10^{10}$\\
\hline
\end{tabular}
%\end{center}
\caption{Estimate of the emitted frequencies, using standard silica fiber parameters: $A_{eff} =10^{-12} \,\, m^2$, $|\beta_2|=10^{-25} \,\, s^2m^{-1}$,  $\beta_1 = 0.5 \times 10^{-8} \,\, sm^{-1}$, $n_2 = 2 \times 10^{-20} \,\, m^2W^{-1}$ and a wavelength $\lambda_0 = 800 \,\,nm$ }
\label{table}
\end{table}
Equation (\ref{omegae}) clearly shows the quantum origin of the predicted effect: It depends on the product of the nonlinearity $n_2$ and the reduced Planck constant $\hbar$. Indeed Eq. (\ref{omegae}) strongly resembles the corresponding result for a BH in Eq. (\ref{HawkT}), in that the emission frequency
is directly proportional to the Hawking temperature: When increasing the soliton ``mass''  (the number of photons $n_0$), or the soliton Schwarzschild time (proportional to $\Delta t_0$), the quantum emission cools and disappears in the classical limit of a large number of photons. We have hence clear evidence that the quantum spreading for soliton is due to black-body like emission. 
Table (\ref{table}) shows the position of the spectral peak according to Eq. (\ref{omegae}) for various pulse durations and parameters for a standard silica optical fiber, and reveals that the emission spectral peak can be tunable in a wide spectral range.

In summary, we have shown that quantum soliton evaporation is directly analogous to black hole evaporation, and may be considered a typical quantum effect for nonlinear waves.
A key feature is that our analysis is based on a fully nonlinear quantized field theory with no resort to linearization.
This gives insights on the way the standard BH evaporation, commonly derived as a linear field theory with
background, can emerge from a future quantum theory of nonlinear gravity. This analogy may stimulate
further developments for quantum gravity inspired by the quantum inverse scattering theory, and quantum solitons.
In addition, the emission spectrum has specific dependency on the soliton parameters such as energy, duration and nonlinearity.
This dependency makes the experimental evidence of soliton evaporation feasible that can be used as a proxy for the observation of black hole evaporation.
The predicted effect is purely quantum as the emitted frequency is directly controlled by the soliton parameters, and may be potentially exploited for realising novel tunable quantum light. 

\acknowledgments{We acknowledge discussions with Maria Chiara Braidotti and Giulia Marcucci. This publication was made possible through the support of a grant
from the John Templeton Foundation (58277). The opinions expressed in this publication are those of the authors and do not necessarily reflect the view
of the John Templeton Foundation.}

%\bibliography{Biblio}

\end{document}